\RequirePackage{fix-cm}
\documentclass[smallextended]{svjour3}       % onecolumn (second format)
\smartqed  % flush right qed marks, e.g. at end of proof
\usepackage{graphicx}
\graphicspath{{./Figs/}}
 \journalname{Journal of Colloids and Polymer Science}
\usepackage{amssymb}
\usepackage{siunitx}
\usepackage{amsmath}

\newcommand{\SImrange}[3]{#1\,\si{#3}-#2\,\si{#3}} %SIrange for mathmode, use: \SImrange{10}{15}{\nano\metre}

\newcommand{\SIpmE}[4]{(#1 \pm #2)\cdot 10^{#3}\,\si{#4}}
\newcommand{\SIE}[3]{#1\cdot 10^{#2}\,\si{#3}}

\newcommand{\abs}[1]{\left| #1 \right|}

\newcommand{\chlamy}[1][{ }]{\textit{Chlamydomonas reinhardtii#1}} % optional argument for punctuation marks

\usepackage{xcolor}

\newcommand{\sub}{\color{gray}}
\definecolor{lightgrey}{rgb}{0.6,0.6,0.6}

\begin{document}

\title{Diffusive dynamics of elongated particles in active colloidal suspensions of motile algae \thanks{Supported by Deutsche Forschungsgemeinschaft Grant ER 467/14-1}
}


\author{Florian von R\"uling         \and
        Francine Kolley  \and
        Alexey Eremin %etc.
}


\institute{Florian von R\"uling \at
              Otto von Guericke University Magdeburg, Institute of Physics, 39106 Magdeburg, Germany
 \\
%              Tel.: +123-45-678910\\
%              Fax: +123-45-678910\\
              \email{florian.rueling@st.ovgu.de}           %  \\
           \and
           Francine Kolley \at
              Otto von Guericke University Magdeburg, Institute of Physics, 39106 Magdeburg, Germany
              Present address: TU Dresden, Center for Advancing Electronics (cfaed), 01069 Dresden, Germany  
              \and
           Alexey Eremin \at
              Otto von Guericke University Magdeburg, Institute of Physics, 39106 Magdeburg, Germany
}

\date{Received: date / Accepted: date}

\maketitle

\begin{abstract}
%The enhancement of the diffusion of elongated silica particles due to interactions with the motile microalgae \textit{Chlamydomonas reinhardtii} was explored in thin capillaries using a particle tracking algorithm and video microscopy. Depending on the number of swimmers, the translational and rotational diffusion constants of sedimented doublets of silica beads can be increased by several orders of magnitude in comparison to purely Brownian motion. For low concentrations of algae, a passive particle exhibits effective Brownian motion in a fluctuating flow field generated by multiple swimmers. Occasional large displacement events that occur when individual swimmers are in the vicinity of the particle or collide with it become frequent upon increasing the concentration of algae which results in strong, diffusive transport.\\
%At a high concentration of \chlamy[,] the algae formed dense clusters at the capillary bottom. Passive particles were enclosed in an active lattice of algae with an active bath above the clusters. The rotational motion of a passive doublet can become subdiffusive in this state of the active colloidal system. The translational motion was diffusive, but with moderate diffusion coefficients despite high concentrations of algae.\\
Swimming microorganisms can influence the diffusion of passive particles. The effect of this swimmer-particle interaction depends on different properties, such as the hydrodynamic field of the swimmer and the relative sizes of microorganisms and particles. We investigated an enhancement of the diffusion of silica doublets in a suspension of microalgae \textit{Chlamydomonas reinhardtii} in a flat capillary. Depending on the concentration of microswimmers, the translational and rotational diffusion constants increase by several orders of magnitude in the presence of the swimming algae. For low concentrations of algae, the doublets exhibit Brownian motion in a fluctuating flow field generated by multiple swimmers. One can observe strong, diffusive transport caused by occasional large displacements. At high swimmer concentration, the algae form dense clusters, where the rotational motion of the doublets shows a subdiffusive behaviour while the translational motion remains diffusive.

%at the bottom of the capillary. The passive particles are than enclosed in in an active lattice of algae with an active bath above the clusters. The rotational motion of doublets can show a subdiffusive behaviour while the translational motion is diffusive.}
\keywords{Active matter \and microswimmers \and diffusion}
\end{abstract}

\section{Introduction}
\label{intro}

Swimming microorganisms like the motile microalgae \chlamy dwell in complex liquid-infused environments, where they encounter suspended or sedimented immersed passive particles. 
The diffusive motion of the particles can be enhanced by self-propelled swimmers due to hydrodynamic interactions~\cite{Leptos:2009,Kurtuldu:2011,Jeanneret:2016} and collisions~\cite{Wu:2000}. Thus, swimming microorganisms can accomplish fluid mixing on the microscale. As a result, small self-propelled objects become crucial for the transport processes in their environment. It was demonstrated, that the flow generated by flagellar beating is necessary for sufficient nutrient uptake and waste removal of certain multicellular organisms~\cite{Short:2006}.  However, the question whether the biogenic contribution to mixing is significant is controversial even for macroscopic swimmers like centimeter-sized krill~\cite{Katija:2012}.\par
The investigation of interactions of microswimmers and passive objects is of great interest from a physicist's perspective.  A fluid containing self-propelled objects is out of thermodynamic equilibrium due to the energy input at the level of the single swimmers~\cite{Yeomans:2016}. Thus, a passive particle immersed in such an active fluid is an experimental model system of nonequilibrium statistical mechanics~\cite{Maggi:2014}. The athermal hydrodynamic and steric contributions to the particle transport can result in intriguing phenomena like the anomalous diffusion of anisometric objects~\cite{Peng:2016}.\par
Furthermore, the employment of swimmer-particle interactions in microfluidic applications is auspicious. Experimental research suggests that microswimmers can serve as actuators~\cite{Sokolov:2010} or, exploiting taxes of biological swimmers, accomplish asymmetric mixing~\cite{Kim:2007}.\par
The enhancement of passive particle diffusion occurs for a variety of swimmer-particle combinations~\cite{Wu:2000,Peng:2016,Yang:2016,Kurtuldu:2011,Jeanneret:2016}. Yet, the details depend on the swimmer  hydrodynamics~\cite{Peng:2016,Yang:2016} and the particle size~\cite{Mathijssen:2018}. In previous experiments the motion of swimmers was confined to quasi-2D~\cite{Kurtuldu:2011,Wu:2000,Peng:2016,Yang:2016}.\par % Wu, Peng, Yang, Kurtuldu 
In this paper, we report an experimental study on the motion of sedimented elongated passive particles in the presence of \chlamy in a three-dimensional fluid.
% Role of particle size (Jeanneret), Role of swimmer hydrodynamics (Peng/Yang)
\section{Materials and Methods}
\label{sec:experimental}
\subsection*{\textbf{{\sub Materials}}}
Wild-type \chlamy of strain SAG 11-32a were purchased from the Culture Collection of Algae at G\"{o}ttingen University. The microalgae were grown axenically in tris-acetate-phosphate medium (TAP) (see Refs.~\cite{Gorman:1965,Hutner:1950}) with air bubbling on a 14h-10h day-night-cycle. The strain was maintained by weekly sub-culturing.\par
Silica beads (MicroSil Microspheres SS06N from Bangs Laboratories, Inc. Mean diameter: \(\SI{4.89}{\mu m}\)) were used as passive particles. Beads were washed in distilled water multiple times using a centrifuge and then stored in a fridge. The spheres were heated to room temperature and redispersed prior to adding them to the suspension of algae. Occasionally, elongated objects consisting of two beads were found (see Fig. \ref{fig:Scheme}c). The maximal axial length of such silica doublets (about \(\SI{10}{\mu m}\)) is comparable to the diameter of \chlamy (\(\SImrange{5}{10}{\mu m}\)). The algae were re-dispersed in distilled water to reduce sticking of the silica beads to the glass.  Polyethyleneglycol (PEG, purchased from Sigma Aldrich, typical \(M_v\) approx. 900'000), was added to the final concentration of \(\SI{0.1}{wt.\%}\) of PEG.\par
\par
Samples were observed in silanized rectangular glass capillaries (Rectangle Boro Tubing VitroTubes from CM Scientific, thickness: \(\SI{500}{\mu m}\), width=\(\SI{5}{mm}\), length: several centimeters). The silanization procedure, which additionally avoids sticking of \textit{Chlamydomonas} to glass,  was based on a method described in Ref.~\cite{Highfield:1983}. 
%The height of the capillary is large enough to allow for three dimensional motion of the swimmers.}\newline

%{\Vorschlag The interaction of \chlamy with silica doublets were observed in a flat capillary (height of $500 \mu m$). The height of the capillary is large enough to allow for three dimensional motion of the swimmers. The motion of a doublet was observed in plane and thus is quasi 2-D. The algae have a diameter of $5-10~ \mu m$. In comparison to this, the maximal axial length of the doublet is approximately $10~ \mu m$. This means, that the size of the algae and the size of the doublet are of same order of magnitude.} 

\subsection*{\textbf{{\sub Methods}}}

\begin{figure}
\centering
\includegraphics[width=12cm]{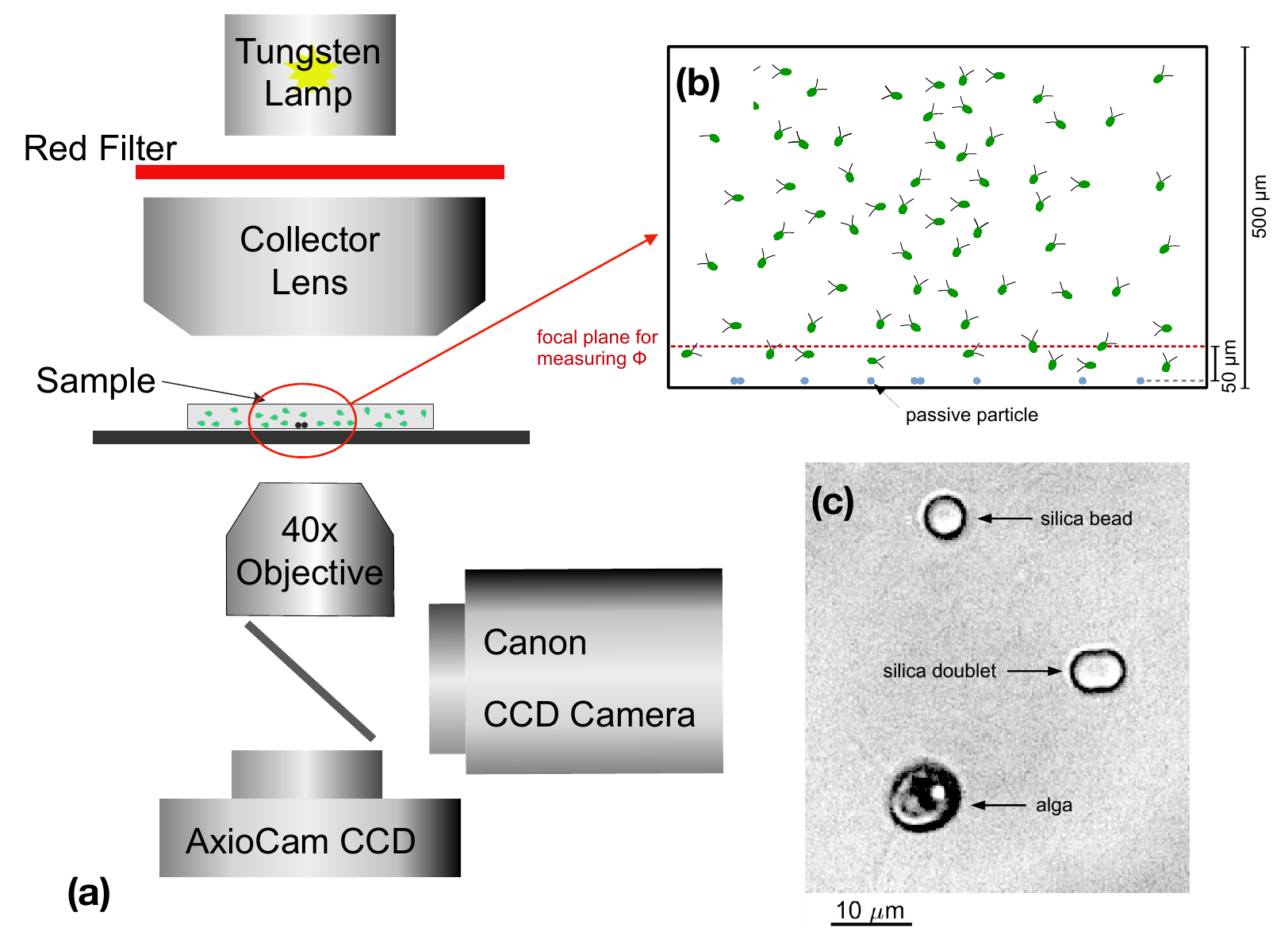}
\caption{Explanation of the experimental setup and conditions. a) Scheme of the experimental setup. b) Schematic side view of a capillary with motile algae and sedimented passive particles. The dashed red line represents the focal plane of videos recorded to measure the area fraction of swimming algae \(\varPhi\). c) Micrograph of a \chlamy[-cell] and passive silica particles.  }
\label{fig:Scheme}
\end{figure}

A sketch of the experimental setup is provided in Fig. \ref{fig:Scheme}a.

Active colloidal suspensions of \chlamy were observed using an inverted microscope (Zeiss Observer.D1).
%Active colloidal suspensions of \chlamy were observed in a flat capllary using an inverted microscope (Zeiss Observer.D1).
Schott KL 2500 LCD cold light source with a red longpass filter (\(\lambda \geq \SI{630}{nm}\)) was used for the ambient illumination of the cells to suppress phototaxis.
The sample was also illuminated under microscope with a halogen lamp (Zeiss HAL 100) through a longpass filter (light intensity \(\approx \SI{30}{W/m^2}\), wavelength \(\lambda \geq \SI{650}{nm}\)). 
%Additionally, there was illumination through a red filter (\(\lambda \geq \SI{630}{nm}\)) from the side with a cold light lamp (Schott KL 2500 LCD) to suppress phototaxis.\\
The cells were kept for \(\SI{30}{min}\) on the microscope stage before data acquisition to allow the algae to adopt to the illumination conditions.%\par
 %Before acquiring video data, the illumination settings were made and samples were left at the microscope stage for about \(\SI{30}{min}\) to allow \chlamy to adapt to the lighting conditions. \\% At high concentration of algae, many cells can gather at the capillary bottom, where the passive particles, being denser than the fluid, lie. The upper capillary wall usually is free from algae. To avoid interactions between passive particles and algae at the capillary bottom that are not swimming, the capillary can be turned upside down after acclimatization. By beginning data acquisition as soon as silica beads are sunken down again, one might then be able to record motion of the passive particles before too many algae have gathered at the bottom. \\
The motion of passive particles was recorded using 40$\times$ magnification (Objective: Zeiss LD Plan-NEOFLUAR 40x/0.6 korr) and AxioCam HRc (Carl Zeiss GmbH) CCD camera ($1-5$ fps  frame rate, and the resolution of 1388$\times$1040). Typical acquisition times were 30 min and 10 min. To investigate silica particles without swimmers, we used the acquisition framerate: \(\SI{0.2}{fps}\) and acquisition time: 90 min.\par
% Video data showing the motion of silica particles in the active bath, acquired at 40x magnification (Objective: Zeiss LD Plan-NEOFLUAR 40x/0,6 korr) using an AxioCam HRc (Carl Zeiss GmbH). The monochrome videos have an acquisition framerate of \(\SI{1}{fps}\) or maximal speed (roughly \(\SI{5}{fps}\)) and a resolution of 1388x1040.  Typical acquisition times were 30 min and 10 min. To investigate silica particles without swimmers, we used acquisition framerate: \(\SI{0.2}{fps}\) and acquisition time: 90 min.\\
%To characterise the active bath interacting with the silica particles, videos of \chlamy close to the lower capillary wall  were acquired
To characterise the fast dynamics of microswimmers interacting with the silica particles, videos of \chlamy in a thin fluid layer located about \(\SI{50}{\mu m}\) above the doublets  (see Fig. \ref{fig:Scheme}b) were acquired at a magnification of 20$\times$ (Objective: Zeiss LD Plan-NEOFLUAR 20$\times$/0.4 corr) with a Canon EOS 600D CCD camera (acquisition framerate: circa \(\SI{60}{fps}\), resolution: 1280$\times$720). The videos were taken before and after acquiring the 40$\times$-videos showing the motion of passive particles.  Using a particle tracking algorithm, the area fraction of swimming algae was extracted from those videos. Area fractions of clustering algae were measured in the same way in the plane of the doublets. When dense clusters of algae occurred, the number of passive objects in this plane can be neglected compared to the cell number.\par
Theoretical values for the diffusion of passive particles  in a thermal bath were calculated for comparison with experimental values using equations for cylindrical particles~\cite{Tirado:1984iv}:
\begin{align}
\label{eq:Dr}
D_{\rm{r}} &= \frac{3 k_B T}{\pi \eta L^3} \cdot \left( \ln p + \delta_\perp\right)\\
\label{eq:Dtpar}
D_\parallel &= \frac{k_B T}{2 \pi \eta L} \cdot \left( \ln p + \nu_\parallel\right)\\
\label{eq:Dtperp}
D_\perp &= \frac{k_B T}{4 \pi \eta L} \cdot \left( \ln p + \nu_\perp\right)
\end{align}
where \(D_{\text{r}}\) is the rotational diffusion coefficient \and \(D_{\parallel ,\perp}\) are  the  translational diffusion constants for motion along the major and minor axis of the particle, respectively, \(L\) is the length and \(p\) is the aspect ratio of the particles which are immersed in a fluid of viscosity \(\eta\) at a temperature \(T\). \(k_B\) is the Boltzmann constant.\(\delta_\perp,\,\nu\) are given in~\cite{Tirado:1984iv}:
\begin{align}
\label{eq:delta1}
\delta_\perp &= -0.662 + \frac{0.917}{p} - \frac{0.05}{p^2}\\
\label{eq:delta2}
\nu_\parallel &= -0.207 + \frac{0.98}{p} - \frac{0.133}{p^2}\\
\label{eq:delta3}
\nu_\perp &= 0.839 + \frac{0.185}{p} + \frac{0.233}{p^2}
\end{align}
%\begin{equation}
%\begin{align}
%\delta_\perp &= -0.662 + \frac{0.917}{p} - \frac{0.05}{p^2}\\
%\nu_\parallel &= -0.207 + \frac{0.98}{p} - \frac{0.133}{p^2}\\
%\nu_\perp &= 0.839 + \frac{0.185}{p} + \frac{0.233}{p^2}\\
%\nu &= \frac{\nu_\parallel + \nu_\perp}{2}
%\end{align}
%\label{eq_delta}  
%\end{equation}
In our system, the motion of passive particles is restricted to two dimensions and the translational diffusion coefficient \(D_\text{t}\) can be obtained using Eqs. \eqref{eq:Dtpar} and \eqref{eq:Dtperp} 
\begin{equation}
\label{eq:Dt}
D_\text{t} = \frac{D_\parallel + D_\perp}{2}
\end{equation} 

Substituting Eqs.~\eqref{eq:delta1}-\eqref{eq:delta3} into Eqs.~\eqref{eq:Dr}-\eqref{eq:Dtperp} and \eqref{eq:Dt} with the rod length \(L=\SI{10}{\mu m}\), aspect ratio \(p=2\), fluid temperature \(T=\SI{293}{K}\) and fluid viscosity \(\eta = \SI{1.76}{mPas}\) (viscosity of distilled water with 0.1 wt.\% PEG measured in~\cite{Kolley:2017}), we obtain: \(D_\text{r}\approx\SIE{1.0}{-3}{rad^2/s}\), \(D_\text{t}\approx\SIE{3.3}{-2}{\mu m^2/s}\). \\

\section{Results and Discussion}

\subsection*{\textbf{{\sub Interaction of freely swimming \chlamy with passive particles}}}

We investigated the motion of silica doublets in an active bath of \chlamy for different swimmer concentrations. As a measure of the concentration of active swimmers, we used the area fraction of moving algae \(\Phi\) measured close to the lower wall of the capillary (approximately \(\SI{50}{\mu m}\) above the wall) rather than the volume fraction.  \(\Phi\) was varied from \(\Phi = \SI{0}{\%}\) to \(\Phi \approx \SI{8.6}{\%}\), the latter corresponds to a cell number density of approximately \(\SIE{1.02}{8}{ml^{-1}}\) or a volume fraction of about \(\SI{5.4}{\%}\).\par
Note that the silica particles, having a higher density than the surrounding medium, sink to the bottom of the glass capillary. Thus, they are located close to a solid boundary. %The diffusive motion of the silica doublets in a thermal bath, i.e. in the absence of microalgae, was slower than predicted by the theory~\cite{Tirado:1984iv}, 
The experimental values for the diffusion coefficients of a doublet in a the absence of swimmers read: \(D_\text{r} = \SIpmE{5.3}{0.3}{-4}{rad^2/s}\), \(D_\text{t} = \SIpmE{5.3}{0.4}{-4}{\mu m^2/s}\).

In order to compare the experimental and the theoretical values for the diffusion constants of passive particles near the wall, the Fax\'{e}n's correction, considering the hydrodynamic interaction with the wall, must be taken into account ~\cite{Leach:2009}. The effect of the solid boundary was studied for a sphere of equivalent volume using the Einstein relation with drag coefficients corrected according to~\cite{Leach:2009}:

\begin{align}
\gamma &= \frac{\gamma _0}{1-\frac{9}{16} \cdot \frac{R}{h} + \frac{1}{8} \cdot \left(\frac{R}{h}\right)^3} \label{eq:Faxen_trans}\\
\beta &= \frac{\beta _0}{1- \frac{1}{8} \cdot \left(\frac{R}{h}\right)^3}\label{eq:Faxen_rot}
\end{align} 
where \(R\) is the radius of the sphere and \(h\) the distance between the centre of the sphere and the wall. For \(h\to R\), the correction yields a change of 44\% for the translational and 13\% for the rotational diffusion constant. This value is too small to explain the discrepancy between measured and predicted values. 
%{\DEL Thus, Fax{\'e}n correction does not change the order of magnitude of the theoretical values for diffusing cylinders. Comparison of experimental diffusion coefficients \(D_\text{r,t}\) with the theoretical values corrected by the Fax{\'e}n factors \(D^{\text{theor.}}_{\text{r,t}}\) yields: \(D^{\text{theor.}}_{\text{r}}/D_\text{r} \approx 1.7\), \(D^{\text{theor.}}_{\text{t}}/D_\text{t} \approx 35\).} 
 %The passive particles are located near the bottom glass plate. Therefore, they are not able to diffuse freely. 
 Therefore, this difference between the experiment and the theory can be attributed to the adhesion of particles to the capillary wall. In addition, the deviation between the  values is much more pronounced for translational than for rotational motion. This can be seen as an indication that the particles are able to rotate easier about the contact area. At the same time, the translational displacement is hindered. 
\begin{figure*}[h!]
\centering
\includegraphics[width=0.8\columnwidth]{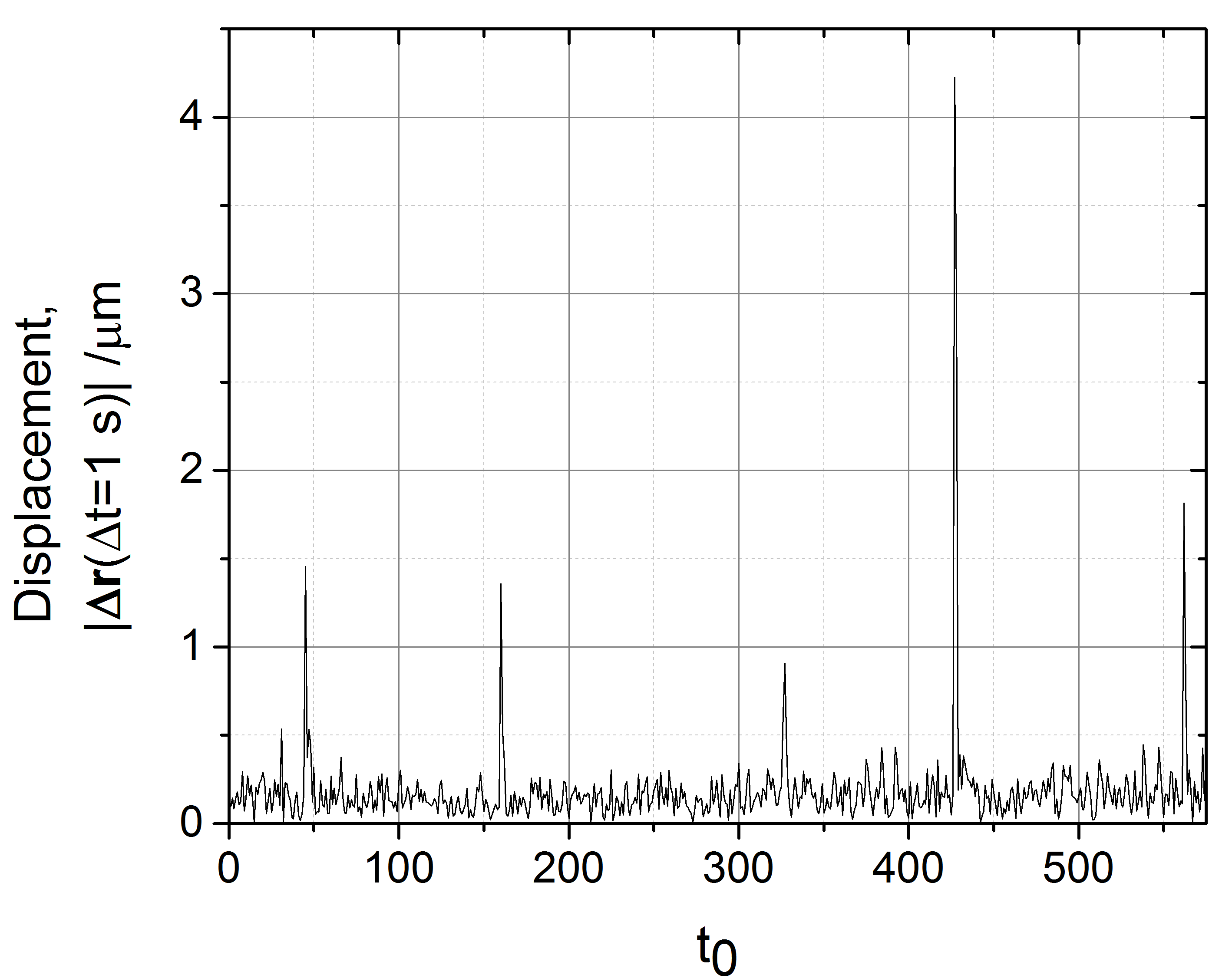}
\caption{Absolute value of the displacement \(\abs{\vec{r} (t_0+\Delta t) - \vec{r} (t_0)}\) of a silica doublet in a dilute suspension of \chlamy (area fraction \(\Phi \approx \SI{0.3}{\%}\)) in the time interval \(\Delta t = \SI{1}{s}\) for different starting points \(t_0\).}
\label{fig:displacement} 
\end{figure*}

The motion of algae significantly enhances the positional and orientational fluctuations of the passive particles. The magnitude of the particle displacement in a fixed time interval is shown in Fig. \ref{fig:displacement} for low swimmer concentration. Long periods of relatively small particle displacements are interrupted by distinct peaks that signal strong translations. Those peaks correspond to either  contact interactions or the hydrodynamic disturbances caused by the swimmers in the close vicinity of the particles. During the periods of small particle displacement, there are no algae in the vicinity of the particle which then is subjected to a collective hydrodynamic field that is the superposition of the far fields the swimmers generate. This flow field is fluctuating since swimmers translate and reorient, thus causing effective Brownian motion of the passive particle. Note that during the periods of small displacement, between the collisions, the diffusive motion is already significantly enhanced with respect to purely thermodynamical diffusion. As the swimmer concentration is increased, the frequency of the large displacement events grows until the periods of small displacement vanish. % at \Phi \approx 0.3: D_rot one order of magnitude larger, D_t two orders of magnitude larger than without swimmers.
Owing to the interactions with swimming algae, both rotational and translational diffusion constants are significantly enhanced in the active bath of \chlamy[.]

\begin{figure}[h!]
\centering
\includegraphics[width=0.8\columnwidth]{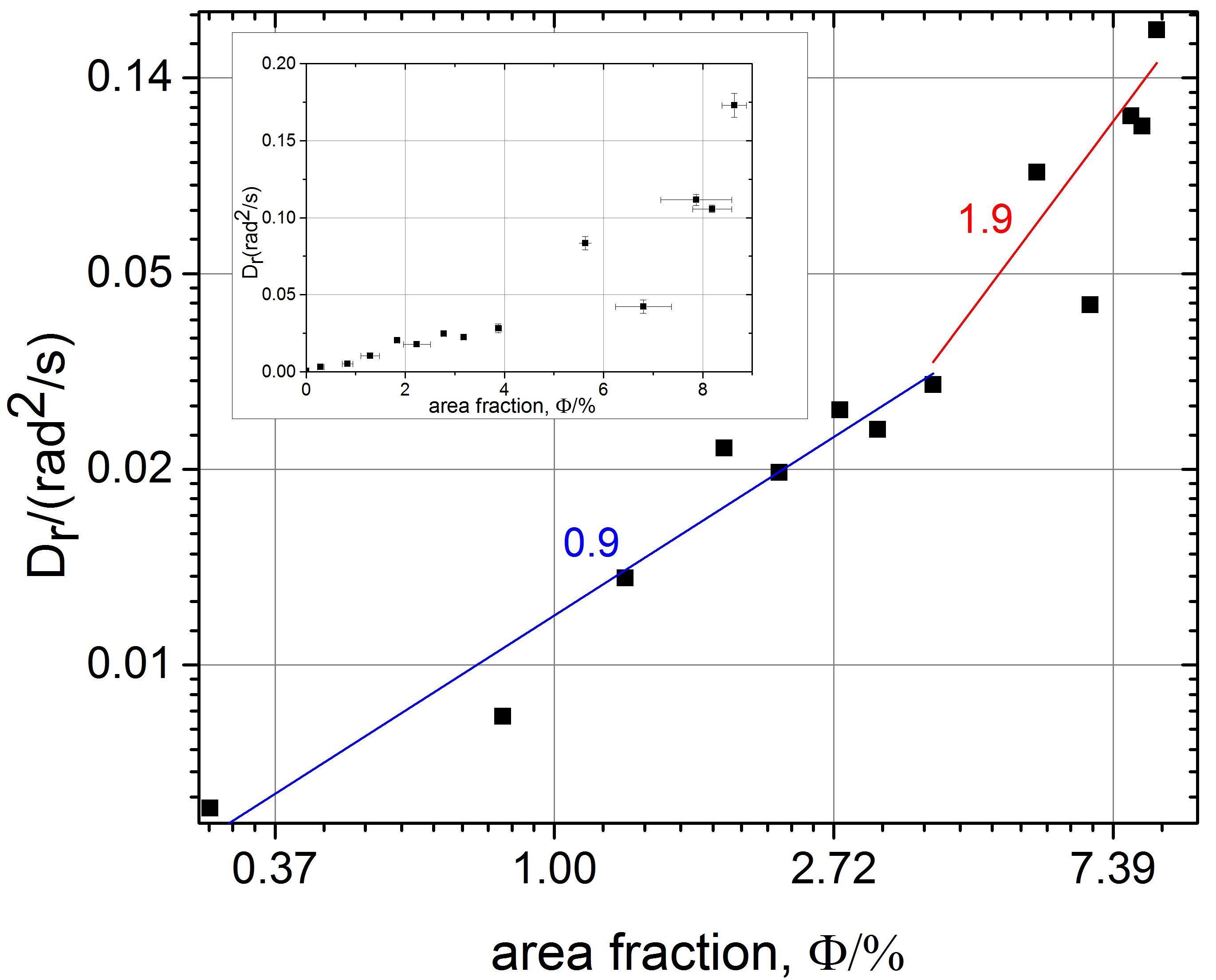}
\caption{The rotational diffusion constant \(D_\text{r}\) of elongated silica particles in active baths with different area fractions  \(\Phi\) of motile \chlamy[.] The blue and red numbers denote the slope of the linear fits in the double-logarithmic scaling for low and high concentrations of algae, respectively. Linear scaling is shown in the inset. }
\label{fig:D_r}
\end{figure}

The rotational diffusion constant \(D_\text{r}(\Phi)\) is shown for different values of the area fraction of moving algae \(\Phi\) in Fig. \ref{fig:D_r}. With increasing concentration of the microswimmers, \(D_\text{r}\) changes by nearly three orders of magnitude. The dependence of \(D_\text{r}(\Phi)\)  is linear for low \(\Phi\), which is in agreement with theory by Pushkin et al.~\cite{Pushkin:2013:a}. Above \(\Phi \approx \SI{3.9}{\%}\) the behaviour becomes nonlinear \(D_\text{r}(\Phi)\propto \Phi^{n}\) with $n> 1.9$. At this volume fraction, the motion of algae remains only weakly correlated. Thus, the crossover at $\SI{3.9}{\%}$ cannot be explained by swarming behaviour as in ~\cite{Peng:2016}. The nonlinearity can arise due to the simultaneous action of the advective flow fields of multiple swimmers on the particle~\cite{Kurtuldu:2011}.
\par
\begin{figure}[h!]
\centering
\includegraphics[width=0.8\columnwidth]{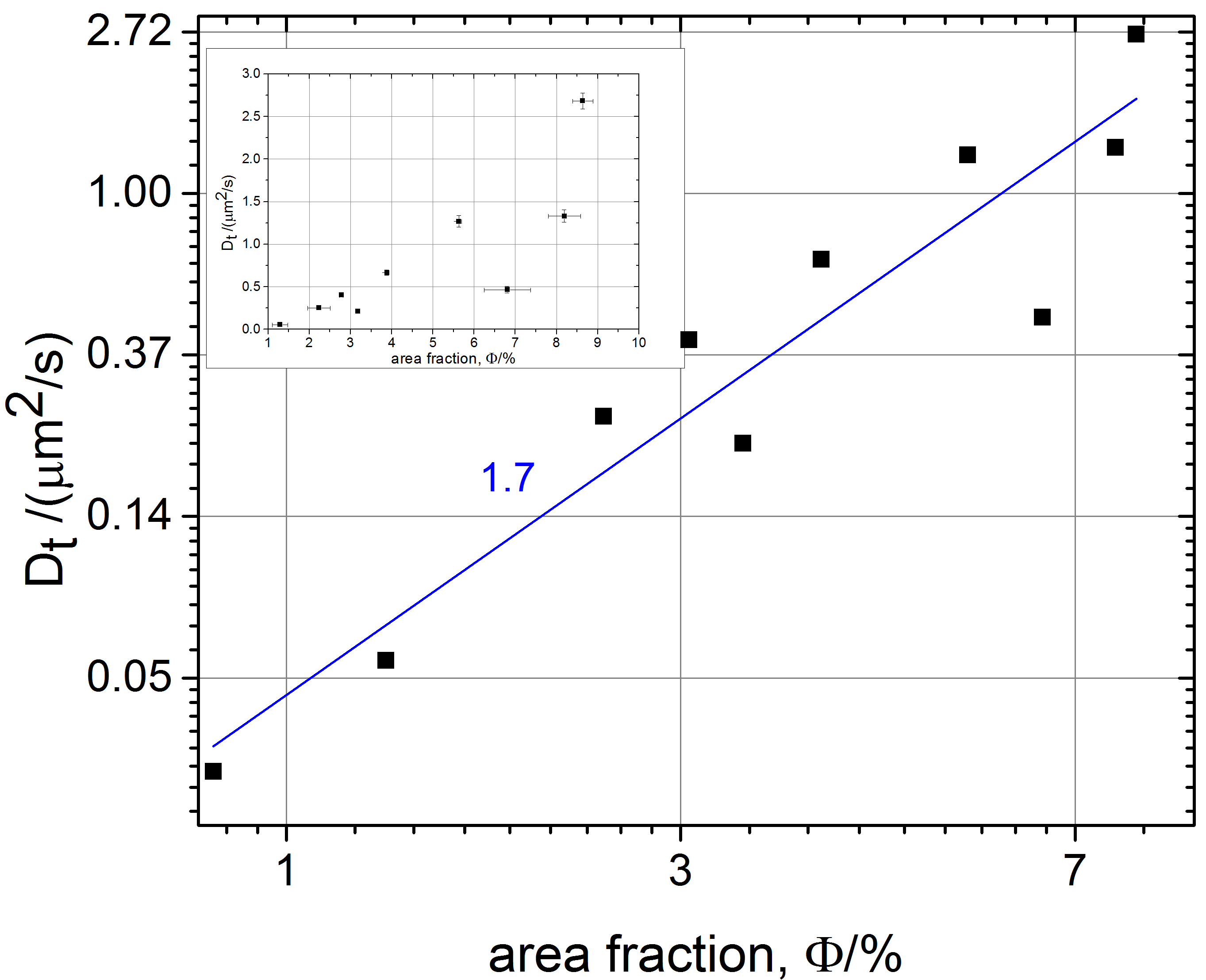}
\caption{The translational diffusion constant \(D_\text{t}\) of elongated silica particles in active baths with different area fractions of motile \chlamy \(\Phi\). The number next to the linear fit in the double-logarithmic plot denotes the slope. Linear scaling is shown in the inset.}
\label{fig:D_t}
\end{figure}

 The translational diffusion constant \(D_\text{t}\) increases by almost four orders of magnitude in the presence of algae as compared to the case of purely thermodynamical diffusion. Intriguingly, the dependence of \(D_\text{t}\) on \(\Phi\) is nonlinear even at low \(\Phi < 3.9\% \) (see Fig. \ref{fig:D_t}). However, \(D_\text{t}\) remains proportional to $\Phi^n $ with $n \approx 1.7$. The unexpectedly strong dependence of the translational diffusion on the number of swimmers is apparently related to the contact adhesion of the passive particles to the capillary wall, which suppresses the linear regime at low area fraction \(\Phi < 3.9\% \). Experiments without swimmers imply that the passive particles can not diffuse freely which is seemingly related to the partial contact of silica particles to the glass wall of the capillary. Yet, algae provide much more energy than the thermally fluctuating fluid and promote particles to overcome the contact adhesion. This situation is different from the rotational diffusion, which is less hindered by the contact to the wall.

%\begin{figure}[h!]
%\centering
%\begin{subfigure}{0.8\columnwidth}
%\centering
%\includegraphics[width=0.8\columnwidth]{D_r}
%\end{subfigure}
%\begin{subfigure}{0.8\columnwidth}
%\centering
%\includegraphics[width=0.8\columnwidth]{D_r_loglog}
%\end{subfigure}
%\caption{The rotational diffusion constant \(D_\text{r}\) of elongated silica particles in active baths with different area fractions \chlamy \(\Phi\) of motile. Top: Linear scaling. Bottom: Double-logarithmic scaling, the blue and red number denote the slope of the respective line.}
%\label{fig:D_r}
%\end{figure}

%\begin{figure}[h!]
%\centering
%\begin{subfigure}{0.8\columnwidth}
%\centering
%\includegraphics[width=0.8\columnwidth]{D_t}
%\end{subfigure}
%\begin{subfigure}{0.8\columnwidth}
%\centering
%\includegraphics[width=0.8\columnwidth]{D_t_loglog}
%\end{subfigure}
%\caption{The translational diffusion constant \(D_\text{t}\) of elongated silica particles in active baths with different area fractions of motile \chlamy \(\Phi\). Top: Linear scaling. Bottom: Double-logarithmic scaling, the blue number denotes the slope of the respective line.}
%\label{fig:D_t}
%\end{figure}

\subsection*{\textbf{{\sub Motion of passive particles enclosed by clusters of algae} }}

\begin{figure}[h!]
\centering
\includegraphics[width=1\columnwidth]{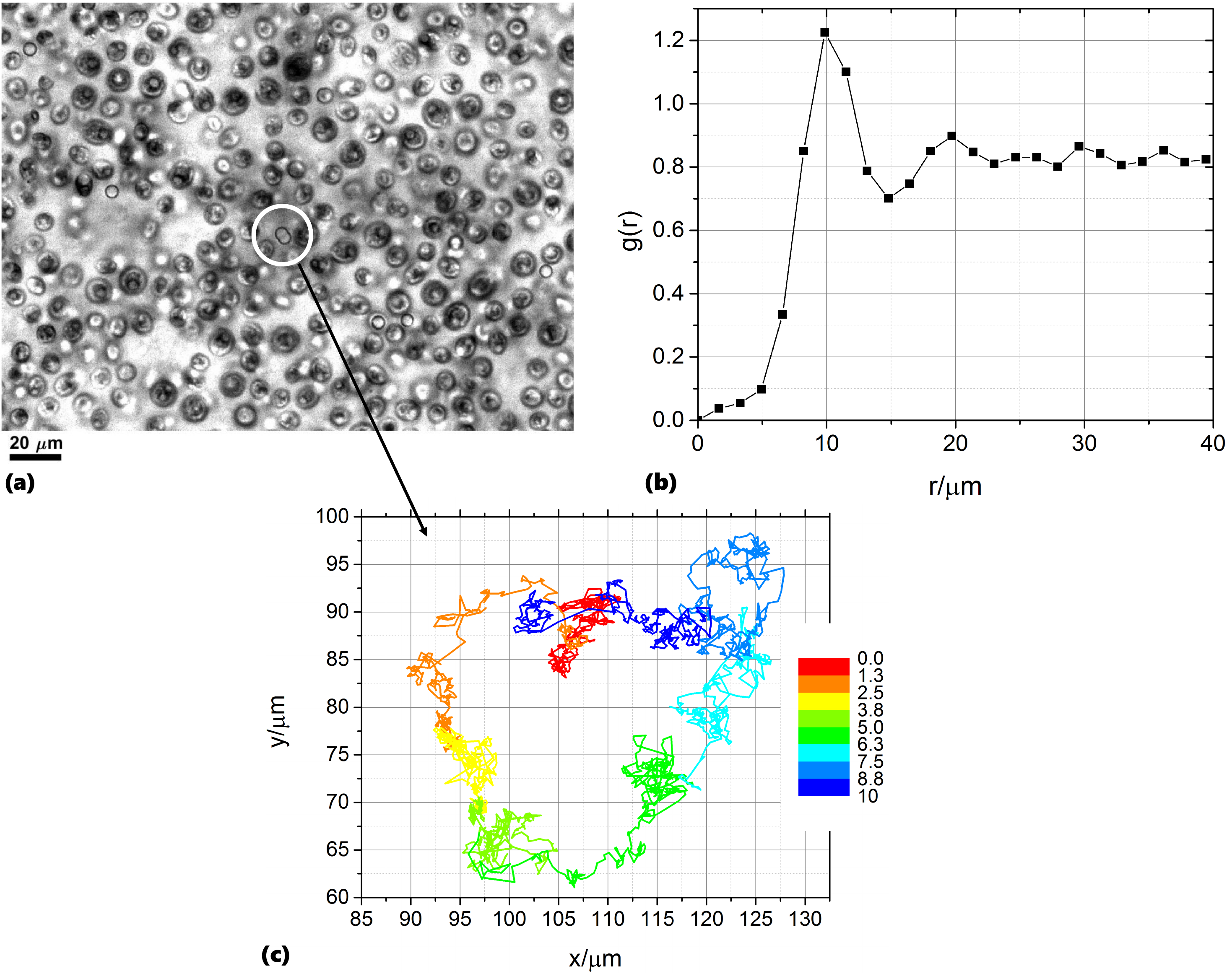}
\caption{(a) At high cell concentrations, \chlamy densely cover the lower capillary wall. Besides the microalgae, few silica particles can be seen. (b) Radial pair correlation function (exemplary) of \chlamy in a clustered state. (c) The trajectory of a silica doublet enclosed by clusters of \chlamy[.] The colour indicates the time in minutes.}
\label{fig:Clustering} 
\end{figure}

Algae tend to adhere to the lower wall of the glass capillary during the experiments. The number of such adhered algae can be strongly reduced by using a suitable illumination in the red range of spectrum~\cite{Baeumchen:2017}. However, at high swimmer concentrations, the number of algae at the surface becomes significant. Fig. \ref{fig:Clustering}a shows an example of such a situation, where the lower capillary wall is densely covered with active but restricted in motion algae which occupy about \(\SI{33}{\%}\) of the area. Positional correlations become stronger pronounced and the motion of algae appears caged. However, in this case the correlations are only short-ranged and short-living (Fig.~\ref{fig:Clustering}b). The motion of passive particles in such a crowded environment becomes obstructed (Fig.~\ref{fig:Clustering}c). Algae can leave the bottom layer or can sediment from the upper layers. Above the clusters of algae a high number of swimmers (comparable to the case with \(\Phi \approx \SI{8.6}{\%}\))  contributes to the active bath.

Due to the sedimentation, the number of algae in the clustered state exhibits a complex dependence on the total number of algae in the capillary. Sparser clusters, covering about \(\SI{15}{\%}\) of the bottom wall area, formed in the case with the largest cell number of \chlamy[.]
%, where also the area fraction of swimming cells took its maximum value \(\Phi \approx\SI{14}{\%}\).
 Note that for the cases with clustering algae the area fractions are only estimates since the accuracy is reduced by the complex background of the videos.% The denoted errors are statistical errors.}%lattice:\(\SIpm{14.6}{1.3}{\%}\);free swimming: \(\Phi =\SIpm{13.9}{0.5}{\%}\), errors are statistical errors

Local ordering of algae can be characterised by the pair correlation function which shows positional correlations of the cells. Since the system is isotropic, it is sufficient to study correlations in dependence on the distance.
The (momentary) radial pair correlation function $g_2(r)=\langle n(r_0) n(r_0+r) \rangle_{r_0}$ of \chlamy[-]cells in the denser clustered state is shown in Fig. \ref{fig:Clustering}b. Here, \(n\) is the number of cells and \(r\) is a radial coordinate. The function $n(r)$ was obtained by cell detection and counting the number of cells in a circular segment of radius $r$ centred around a selected cell (see Ref.\cite{Chaikin}). The pair correlation function has pronounced first and second maxima indicating short correlations between the algae forming small clusters. The neigbouring cells appear often in contact.\par
\begin{figure}
\centering
\includegraphics[width=\columnwidth]{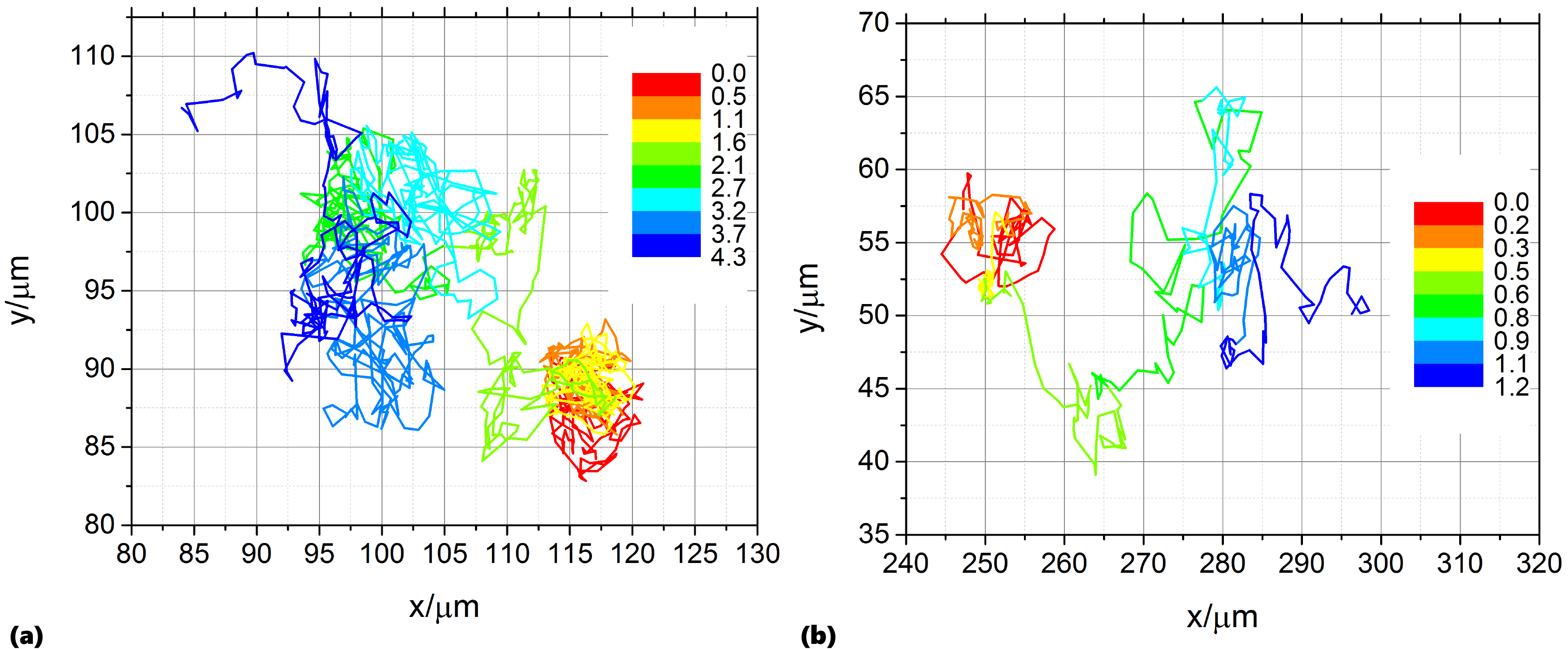}
\caption{Trajectory of a \chlamy[-]alga in the (a) denser and (b) sparser clustered state. The colour coding indicates the time in minutes.}
\label{fig:Traj_alga}
\end{figure}

  %The maxima of $g_2(r)$ appear at multitudes of the typical diameter of algae which implies that neighbouring cells are often in contact.
%The active lattice is dynamic, and the algae at the wall move, although it is not clear whether this happens predominantly due to self-propulsion or due to interactions with freely swimming algae. {\ToDo this is not good. If they are stuck, they are not active any more.} 
Figure \ref{fig:Traj_alga}a and b show the trajectories of algae in the denser and sparser clustered state, respectively. The trajectory in the dense state terminates when the alga swims away. A cell is confined for minutes and can traverse the clusters of algae only slowly, its trajectory reaching an extension of about \(\SI{40}{\mu m}\) in more than four minutes. The path of the alga in the sparser clustered state with a strong active bath (Fig. \ref{fig:Traj_alga}b) spans more than \(\SI{50}{\mu m}\), five times the body length of \chlamy[,] in about 70 seconds. If the algae were adhered to the wall with their about \(\SI{12}{\mu m}\)~\cite{Drescher:2010} long flagella, their centre of mass motion would be restricted to a circle with diameter \(\SI{34}{\mu m}\) under the best conditions. Practically, if the clustering algae  were adhered, this distance would be reduced due to the entanglement of flagella. Thus, algae in the clustered state are not adhered to the capillary wall. The velocities of the clustering algae are one order of magnitude smaller than bulk self-propulsion velocities. A mean squared displacement analysis does not evince any ballistic regime. Yet, the time resolution of the experiments was only about 5 fps and the motion of algae in the clustered state can be strongly randomised due to interactions with the fast algae of the active bath. For an alga in the sparser clustered state, the translational diffusion coefficient \(D_\text{t}^{\text{Alga}}\approx \SI{3.6}{\mu m^2/s}\) can be determined, which exceeds  the translational diffusion coefficients measured for the silica doublets of similar size.
 % estimate: two times length of flagella + 2* body radius (COM-position) = 2*12+2*5=34. True only if flagella are adhered in one point

A passive particle enclosed by clustering algae is driven due to the interactions with algae from the bath and the clusters. However, the clusters also provide a dynamic geometrical constraint: Algae can jam the particle and even form cages around it. The trajectory of a silica doublet surrounded by algae in the dense clustered state is displayed in Fig. \ref{fig:Clustering}c and exhibits intermittent confinement. The translational diffusion constant of this particle  takes only a moderate value of \(D_\text{t} \approx \SI{0.5}{\mu m^2/s}\) and its rotational motion even becomes subdiffusive on a time scale of \(\SI{10}{s}\). Since algae and passive particles are of comparable size, a single alga in the vicinity of the silica doublet is sufficient to hinder its rotational motion significantly.\\

For the sparser clustered state, the translational diffusion constant of the passive doublet \(D_\text{t} \approx \SI{2.1}{\mu m^2/s}\) is smaller than the maximum value despite the very high number of swimming algae. The rotational diffusion constant \(D_\text{r}\approx \SI{0.24}{rad^2/s}\), in contrast, reaches its maximum among the observed values.
\begin{figure}
\centering
\includegraphics[width=0.8\columnwidth]{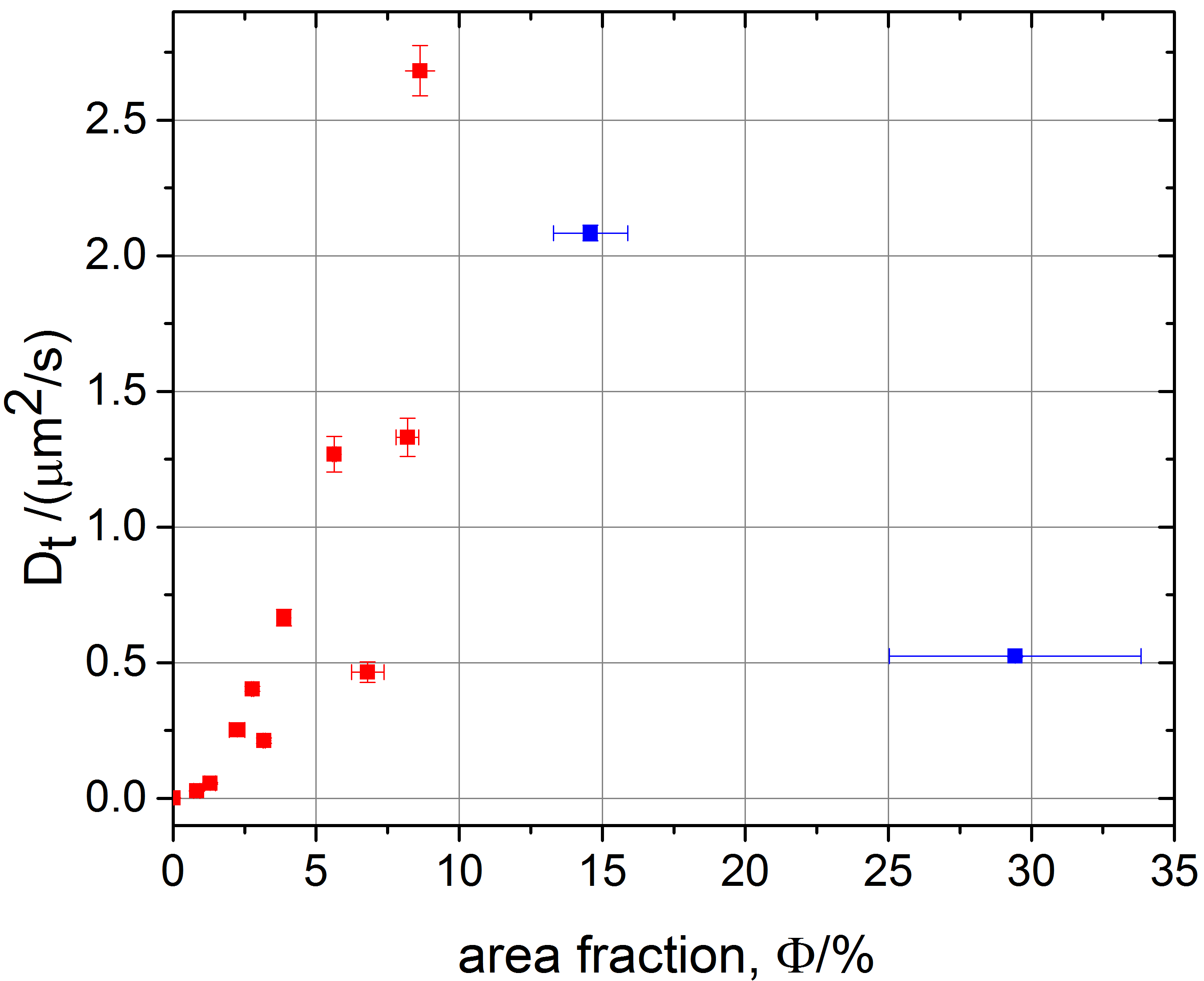}
\caption{The translational diffusion constant \(D_\text{t}\) of elongated silica particles in pure active baths (red symbols) or in active baths with clustering \chlamy (blue symbol) for different values of the area fraction \(\Phi\) of algae participating in the dominant interaction with the passive objects.  For red symbols, \(\Phi\) is the area fraction of motile \chlamy[,] whereas for the blue symbol \(\Phi\) is the area fraction of clustering algae.}
\label{fig:D_t_lat}
\end{figure}
The situations of a particle in a pure active bath or a clustered state with active bath can not be compared easily. The latter case is not well characterised by the area fraction of moving algae. Instead, the area fraction of clustering algae may be a better measure, since the dominating interaction appears to be with the sedimented algae rather than the freely swimming algae. It might still be instructive to combine the data for the translational diffusion constant for both cases. Fig. \ref{fig:D_t_lat} shows the translational diffusion constants for different area fractions of those algae that are involved in the dominant interaction, i.e. either freely swimming or clustering algae. Note that the mere cell concentrations of the data points for the dense clustered state and for the maximum enhancement of diffusion by freely swimming algae are comparable and the data point for the sparser clustered state corresponds to the highest concentration of algae. The translational diffusion constant of the sedimented passive particles does not increase monotonously with the concentration of \chlamy due to the formation of clusters of algae at the lower capillary wall. 
\section{Conclusions}

We demonstrated that the motile microalgae \chlamy can enhance the rotational and translational diffusion constants of sedimented elongated silica particles by several orders of magnitude. The concentration dependence of the diffusion constants of the passive particles exhibits a non-monotonic behaviour due to the crowding of sedimented cells at the solid boundary. Crowding of the cells restricts the motion of the passive particles by caging the particles.

In the high-concentration regime, we observe strong short-range positional correlations of the algae forming small clusters. \chlamy were observed to switch between the freely swimming and the clustered state. In both states the motile algae can affect the motion of a passive particle, yet algae in the clustered state also act as a dynamic geometrical restriction. %A passive particle enclosed in an active lattice with an active bath above it experiences complex interactions.

\begin{acknowledgements}
The authors acknowledge Prof. Claus-Dieter Ohl and Prof. Ralf Stannarius for fruitful discussions. We thank Dr. Hajnalka N\'adasi for assistance in surface treatment and discussions, Dr. Dmitri Puzyrev for assistance with the particle tracking algorithms, Wieland Ivo Schifferm\"uller for participation in the experiments in the frame of the MINT-Praktikum. F.v.R. acknowledges support by a Landesstipendium Sachsen-Anhalt. The research was partially supported by Deutsche Forschungsgemeinschaft (Project ER 467/14-1). We acknowledge technical support by K. Guttmann and Dr. P. Pfeiffer.
\end{acknowledgements}

 \section*{Conflict of interest}

 The authors declare that they have no conflict of interest.

\bibliographystyle{spphys}       % APS-like style for physics
\bibliography{papers.bib}

%\begin{thebibliography}{}
%%
%% and use \bibitem to create references. Consult the Instructions
%% for authors for reference list style.
%\end{thebibliography}

\end{document}